# Exotic properties and optimal control of quantum heat engine

Congjie Ou[1] and Sumiyoshi Abe[1,2,3]

[1]*College of Information Science and Engineering, Huaqiao University, Xiamen 361021, China*

[2]*Department of Physical Engineering, Mie University, Mie 514-8507, Japan*

[3]*Institute of Physics, Kazan Federal University, Kazan 420008, Russia*

**Abstract** — A quantum heat engine of a specific type is studied. This engine contains a single particle confined in the infinite square well potential with variable width and consists of three processes: the *isoenergetic process* (which has no classical analogs) as well as the isothermal and adiabatic processes. It is found that the engine possesses exotic properties in its performance. The efficiency takes the maximum value when the expansion ratio of the engine is appropriately set, and, in addition, the lower the temperature is, the higher the maximum efficiency becomes, highlighting aspects of the influence of quantum effects on thermodynamics. A comment is also made on the relevance of this engine to that of Carnot.

It is well known that thermodynamics has played a crucial role for Planck in framing the quantum hypothesis. However, as pointed out in Ref. [1], it seems that after the formation of quantum mechanics these two fields have developed rather separately, although a few exceptions have appeared in the middle of the last century [2,3]. Today, such a situation is rapidly changing due to developments of quantum state engineering, nanoscience/nanotechnology, and quantum information. Such a new stream would have been initiated in the beginning of this century by the investigations, e.g., in Refs. [4-8]. Ever since, even limited to quantum heat engines, a large number of works have been done (see Refs. [9-21], for example). Also, recent recognition of an intrinsic similarity between quantum mechanics and thermodynamics [22-26] has shed new light on the relevant issues. In addition, progress of thermodynamics of small systems, or commonly referred to as nanothermodynamics, should be noted [27-32]. A point of importance is that it is possible to consider the "smallest" engine containing only a single particle. A recent experimental realization [33] of a heat engine with a single colloidal particle subject to an optical laser trap shows that in fact it makes sense to discuss such an extreme case, although the underlying dynamics there is not quantum but stochastic.

In this paper, we study a specific-type quantum heat engine with a single particle, which shows how quantum effects give rise to a striking phenomenon that is unexpected from the viewpoint of classical thermodynamics. The cycle is reversible and consists of three processes [34], not four. Two of them are familiar ones: the isothermal and adiabatic processes, whereas the other is the *isoenergetic process* [22-26] that does not have its classical counterpart, as discussed below. We report a remarkable finding that the efficiency of this engine depends on the way of expansion and compression and has its unique maximum for each value of the temperature. In other words, the quantum effects make optimal control of the engine possible. In addition, the lower the temperature is, the higher the maximum efficiency becomes, again indicating importance of quantum effects. These results highlight how quantum effects change the classical view of thermodynamics.

Let us start our discussion with describing the quantum heat engine considered here. It consists of a single particle with mass, $m$, confined in a one-dimensional infinite square well potential with width, $L$, that varies slowly. (Although generalization to higher dimensions is straightforward, we shall work with the one-dimensional case since we wish to simplify the situation as much as possible.) There exists only one heat bath, from which the engine receives heat. There is also an external system that controls both $L$ and quantum states, simultaneously, i.e., essentially the combination of mechanical expansion/compression and quantum state engineering, as well as the environmental system that absorbs the heat of the engine. Therefore, it has no analogs in classical theory. The cycle is realized by three processes, as shown in the $f$-$L$ plane in Fig. 1 that is the $P$-$V$ plane in one dimension, where the pressure becomes the force, $f$. The process (I) from the initial state $A$ to $B$ is isothermal, being in contact with the heat bath. The process (II), $B \rightarrow C$, is the isoenergetic one associated with the energy bath, where the internal energy is kept constant. Finally, the process (III), $C \rightarrow A$, is adiabatic. A key is in the isoenergetic process (II), since it is inherently quantum.

All of the processes mentioned above are assumed to be slow. That is, their time scales are much larger than the thermal relaxation time as well as the quantum mechanical one, $\sim \hbar/E$, where $E$ stands for a typical value of the system energy. This makes the cycle reversible, that is, the total change of the (internal) entropy vanishes after one cycle. It also enables one to describe the system states in terms of instantaneous energy eigenstates, $\{|u_n\rangle\}_n$, in the adiabatic approximation. Therefore, we consider the stationary Schrödinger equation for each value of $L$ at each time: $H|u_n\rangle = E_n|u_n\rangle$, where $H$ is the Hamiltonian of the particle confined in the infinite square well potential with instantaneous width, $L$, and $E_n = \pi^2 \hbar^2 n^2/(2mL^2)$ ($n = 1, 2, 3, ...$) are the energy eigenvalues. In contact with the heat bath at temperature, $T$, the density matrix reads $\rho = Z^{-1} \exp[-H/(k_B T)] = Z^{-1} \sum_{n=1}^{\infty} \exp[-E_n/(k_B T)]$ $\times |u_n\rangle\langle u_n|$, where $k_B$ is the Boltzmann constant. $Z = \text{Tr}\exp[-H/(k_B T)]$ is the partition function:

$$Z = \sum_{n=1}^{\infty} \exp\left(-\frac{1}{k_B T}\frac{\pi^2 \hbar^2 n^2}{2mL^2}\right)$$

$$= \frac{1}{2}[\vartheta_3(0, q) - 1], \tag{1}$$

where $\vartheta_3(z, q)$ is the theta function [35] defined by

$$\vartheta_3(z, q) = \sum_{n=-\infty}^{\infty} q^{n^2} e^{2inz} \tag{2}$$

with

$$q = \exp\left(-\frac{1}{k_B T}\frac{\pi^2 \hbar^2}{2mL^2}\right) \tag{3}$$

in the present case.

For the discussion, here it is convenient to calculate the internal energy and force (i.e., the one-dimensional analog of the pressure). The internal energy, $U$, is defined by $U = \text{Tr}(H\rho)$. Its change along a certain process, $dU$, leads to the general formulation of the first law of thermodynamics: $d'Q = dU + d'W$, where the changes of heat and work are given by $d'Q = \text{Tr}(H\,d\rho)$ and $d'W = -\text{Tr}(\rho\,dH)$, respectively. $U$ itself is found to be

$$U = \frac{\pi^2 \hbar^2}{2mL^2}\frac{q\,\dot{\vartheta}_3(0,q)}{\vartheta_3(0,q)-1}, \tag{4}$$

where $\dot{\vartheta}_3(0,q) \equiv d\vartheta_3(0,q)/dq$. The force, $f$, is related to the work as $d'W = f\,dL$, yielding

$$f = -\frac{1}{Z}\sum_{n=1}^{\infty}\frac{\partial E_n}{\partial L}\exp\left(-\frac{E_n}{k_B T}\right) = \frac{2U}{L}. \tag{5}$$

Eqs. (4) and (5) are the essential relations for the subsequent discussions.

To realize the cycle in Fig. 1, it is necessary to analyze the slopes, $\partial f/\partial L$, of each process. The slope of the curve, $C \to B$, is found to be negative: $U = U_B = U_C$ is constant during this process, implying $\partial f/\partial L|_{C\to B} = -2U/L^2 < 0$. Also, the slope of the curve, $A \to C$, is clearly negative, since $TL^2$ is kept constant during the adiabatic process (i.e., the state remains unchanged), and accordingly $\partial f/\partial L|_{A\to C} = -6U/L^2 < 0$. On the other hand, the slope of the curve, $A \to B$, is calculated to be $\partial f/\partial L|_{A\to B} = -6U/L^2 + 4(\Delta H)_I^2/(k_B T L^2)$, where $T$ is fixed, and $(\Delta H)_I^2 = \mathrm{Tr}(H^2 \rho) - U^2$ is the variance of the energy at each state during the process (I), $A \to B$. It does not seem simple to show the inequalities, $\partial f/\partial L|_{C\to B} > \partial f/\partial L|_{A\to B} > \partial f/\partial L|_{A\to C}$, analytically. So, we have performed numerical evaluations and have ascertained that these inequalities, in fact, hold. Consequently, the closed cycle $A \to B \to C \to A$ in Fig. 1 is realizable. Also, after a cycle, the engine returns to the original state $A$, since there is no multivaluedness in the $f$-$L$ plane.

Now, let us analyze the properties of the three processes I-III.

(I) The isothermal process, $A \to B$. The temperature is fixed during this process. Clearly, $T = T_A = T_B$, where $T_A$ ($T_B$) is the value of the temperature at $A$ ($B$). The work is given in terms of the free energy difference

$$W_{A\to B} = k_B T_A \ln \frac{\vartheta_3(0, q_B) - 1}{\vartheta_3(0, q_A) - 1}$$

$$= Q - (U_B - U_A), \tag{6}$$

where $q_A$ ($q_B$) and $U_A$ ($U_B$) are the values of $q$ in Eq. (3) and the internal energy at $A$ ($B$), respectively. Here and hereafter, we use $T_A$ for convenience for the later discussion. $Q$ is the amount of the quantity of heat absorbed from the heat bath during the process. Since the process is isothermal, it is given in terms of the change of the entropy, $Q = T_A (S_B - S_A)$, which is rewritten as follows:

$$Q = \frac{\pi^2 \hbar^2}{2 m L_B^2} \frac{q_B \, \dot{\vartheta}_3(0, q_B)}{\vartheta_3(0, q_B) - 1} - \frac{\pi^2 \hbar^2}{2 m L_A^2} \frac{q_A \, \dot{\vartheta}_3(0, q_A)}{\vartheta_3(0, q_A) - 1} + k_B T_A \ln \frac{\vartheta_3(0, q_B) - 1}{\vartheta_3(0, q_A) - 1}, \qquad (7)$$

where the expression of the free energy, $F = -k_B T \ln Z$, as well as Eq. (4) have been used, and $L_A$ ($L_B$) is the potential width at $A$ ($B$).

(II) The isoenergetic process, $B \to C$. This is a key process in the present discussion. Here, the internal energy is kept constant: $U = U_B = U_C$. It is of importance to note that, in marked contrast to classical thermodynamics, this does not mean that the process is isothermal, since the law of equipartition of energy is violated in the quantum regime. The process is characterized by $dU = 0$, which leads to $d'Q = d'W$. In other words, the effect of change of the potential width is compensated by that of change of the temperature in such a way that the internal energy remains constant. Below, we see that the work is finite, and therefore $d'Q \neq 0$, implying that the engine is in contact with an environment (i.e., an external system) in a way that has no classical counterpart. From the condition, $0 = dU = (\partial U / \partial T)_L dT + (\partial U / \partial L)_T dL$, it follows that

$$\frac{dT}{dL} = -\frac{1}{C_L} \left( \frac{\partial U}{\partial L} \right)_T, \qquad (8)$$

where $C_L = (\partial U / \partial T)_L$ denotes the specific heat at constant volume (i.e., width) given by

$$C_L = \frac{1}{k_B T^2} (\Delta H)_{\text{II}}^2 \tag{9}$$

with $(\Delta H)_{\text{II}}^2$ being the variance of the energy at each state during this process. On the other hand, $(\partial U / \partial L)_T = 2(\Delta H)_{\text{II}}^2 / (k_B T L) - 2 U_B / L$, giving rise to

$$\frac{dT}{dL} = -\frac{2T}{L} + \frac{2 k_B T^2}{L} \frac{U_B}{(\Delta H)_{\text{II}}^2}. \tag{10}$$

For the later purpose, here we use $U_B$. This equation determines how the temperature changes during the process. For a practical purpose, the isoenergetic condition, $U(T, L) = U(T_B, L_B)$ with $T_B = T_A$, may be convenient for evaluation of the temperature of the engine with respect to the width. In Fig. 2, we present the plots of the function, $T = T(L)$, for some values of $T_A$. It is interesting to see an exotic property that the temperature *decreases* for this specific compression. Now, the work is calculated as follows:

$$W_{B \to C} = \int_B^C dL\, f$$

$$= 2 U_B \ln \frac{L_C}{L_B}, \tag{11}$$

which is negative, where $L_C$ is the potential width at $C$.

(III) The adiabatic process, $C \to A$. During this process, $d'Q = 0$, that is, the quantum state is fixed. Therefore, $TL^2$ is kept constant, since the density matrix is the function of the variables in this combination. In particular, holds

$$L_C = L_A \sqrt{\frac{T_A}{T_C}}, \qquad (12)$$

where $T_C$ is the value of the temperature at $C$. The work is

$$W_{C \to A} = U_C - U_A, \qquad (13)$$

which directly comes from the first law of thermodynamics.

From Eqs. (6), (11), and (13), the total work done during a single cycle is found to be $W_{tot} = W_{A \to B} + W_{B \to C} + W_{C \to A} = Q - 2U_B \ln(L_B / L_C)$, where $U_B = U_C$ has been used. Therefore, the efficiency of the engine is obtained as follows:

$$\eta = \frac{W_{tot}}{Q} = 1 - \frac{2U_B}{Q} \ln \frac{L_B}{L_C} \qquad (14)$$

with $Q$ in Eq. (7). Fixing the initial condition on the pair of the temperature and width at $A$ and eliminating $L_C$ by employing Eq. (12), we can express the efficiency as the function of $L_B$ and $T_A$, although such a procedure requires a numerical analysis of Eq. (10). Consequently, given the initial condition, we have the efficiency of the engine as the function only of the expansion ratio $L_B / L_A$. It is of importance to note that this expansion-ratio dependence has its origin in the quantum effects, since it is indivisibly connected to the isoenergetic process (II) with no classical analogs and violation of the law of equipartition of energy.

In Fig. 3, we present the plots of the efficiency with respect to $L_B$ for several values of the temperature, $T_A$, that are numerically calculated using Eq. (10). There, one observes a remarkable behavior. There always exists the specific value of $L_B$ that maximizes the efficiency. In addition, as shown in Table I, the lower the temperature is, the higher the maximum value of the efficiency becomes. (Although increase of the maximum value seems to become slower with decreasing temperature, no saturation is observed.) These are the main results of the present work, which are without analogs in classical thermodynamics.

As mentioned earlier, the engine is reversible. In fact, we have numerically ascertained that the procedure employed here for identifying the process (II) makes the entropy change, $S_B - S_A$ [see the discussion above Eq. (7)], be canceled by $S_C - S_B$ (recall that $S_A - S_C$ vanishes since the process (III) is adiabatic), provided that the infinitesimal change of the entropy during the slow processes is given by $dS = d'Q/T$, which is also identical with the derivative of the von Neumann entropy, $S[\rho] = -k_B \text{Tr}(\rho \ln \rho)$.

Furthermore, a comment is also made on the relevance of the present three-process engine to the Carnot cycle. Consider a pair of three-process engines of the present type with the hot and cold heat baths and combine them in such a way that the total system consists of four processes (i.e., two isothermal and two adiabatic processes) and two isoenergetic processes cancel each other. The resulting efficiency is found to be the celebrated one of Carnot [36]. This may be due to the fact that the heat baths considered here are classical.

In conclusion, we have studied a quantum heat engine of a specific type that consists not only of the isothermal and adiabatic processes but also of the isoenergetic process, which is inherently quantum and does not have its counterpart in classical thermodynamics. We have obtained a remarkable result that the efficiency can be maximized by controlling the expansion ratio. In particular, the lower the temperature is, the higher the maximum efficiency becomes. These highlight in a peculiar manner how quantum effects may change our classical view of thermodynamics.

* * *

The work of CO was supported by the grants from Fujian Province (No. 2015J01016, No. JA12001, No. 2014FJ-NCET-ZR04) and from Huaqiao University (No. ZQN-PY114). He also thanks for Toka-Donghua Educational and Cultural Exchange Foundation for providing him with a scholarship and Mie University for the hospitality extended to him. SA would like to acknowledge the High-End Foreign Expert Program of China for support and the warm hospitality of Huaqiao University. His work was also supported in part by a Grant-in-Aid for Scientific Research from the Japan Society for the Promotion of Science and by the Ministry of Education and Science of the Russian Federation (the program of competitive growth of Kazan Federal University).

# Figure and Table Captions

Fig. 1:   The cycle of the quantum heat engine consisting of three processes depicted in the plane of the force ($f$) and potential width ($L$). The processes $A \rightarrow B$, $B \rightarrow C$, and $C \rightarrow A$ are isothermal, isoenergetic, and adiabatic, respectively.

Fig. 2:   The plots of the temperature with respect to the width along the isoenergetic process for three different values of the temperature in the initial state $A$. Both the energy eigenvalues, $E_n = \pi^2 \hbar^2 n^2 / (2mL^2)$, and $k_B T$ are measured in the unit of the ground-state energy at $A$: $E_{1,A} = \pi^2 \hbar^2 / (2mL_A^2)$. For the sake of simplicity, $\pi^2 \hbar^2 / (2m)$, $k_B$, and $L_A$ are all set equal to unity. Accordingly, both $T$ and $L \, (= L/L_A)$ are dimensionless. Here, the maximal potential width, $L_B$, is chosen to be $L_B = 4$.

Fig. 3:   The plots of the efficiency with respect to the expansion ratio, $L_B \, (= L_B/L_A$ with $L_A$ being unity), for five different values of the temperature in the initial state $A$. The lower panel shows the detailed behaviors of the efficiency around the maxima. All quantities are dimensionless, as in Fig. 2.

Table 1:   The maximum efficiency at different temperatures and the corresponding values of the potential width, $L_B^*$. Note that $L_B^*$ is nonmonotonic with respect to temperature. All quantities are dimensionless.

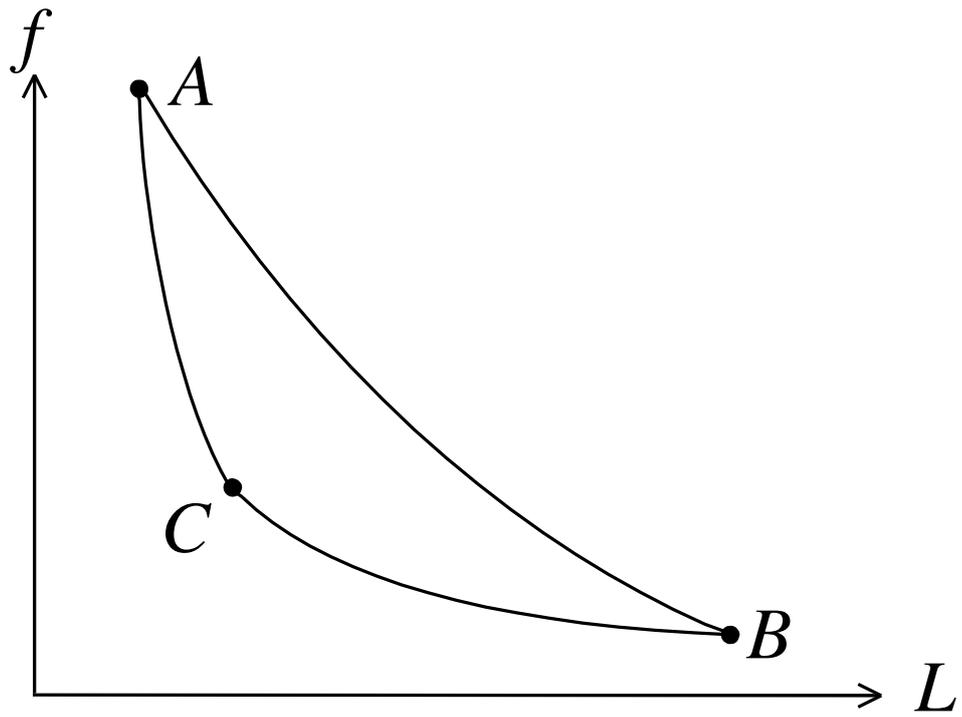

Fig. 1

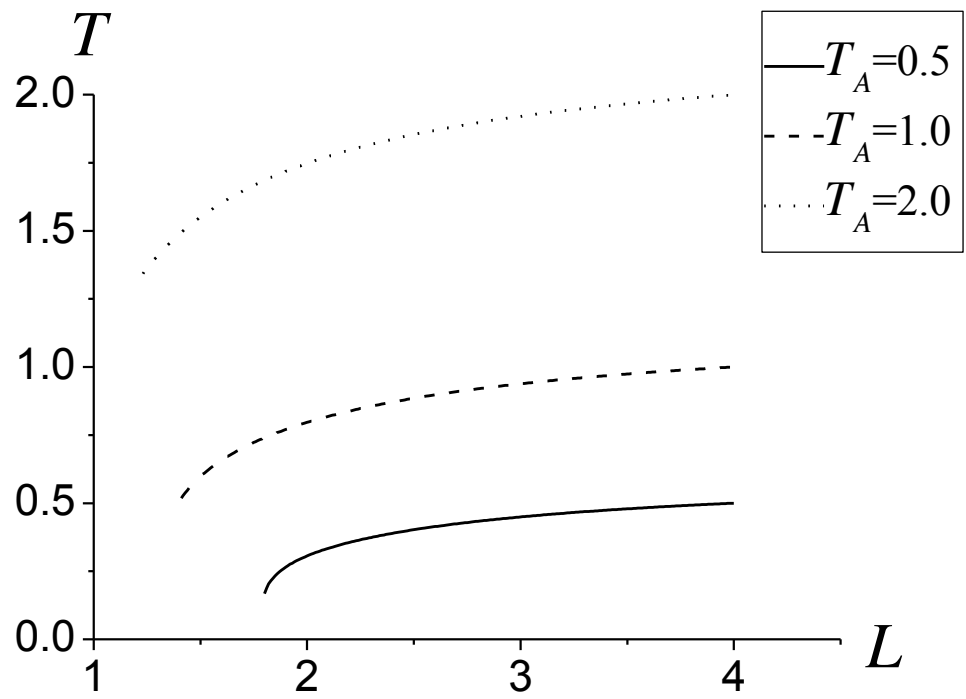

Fig. 2

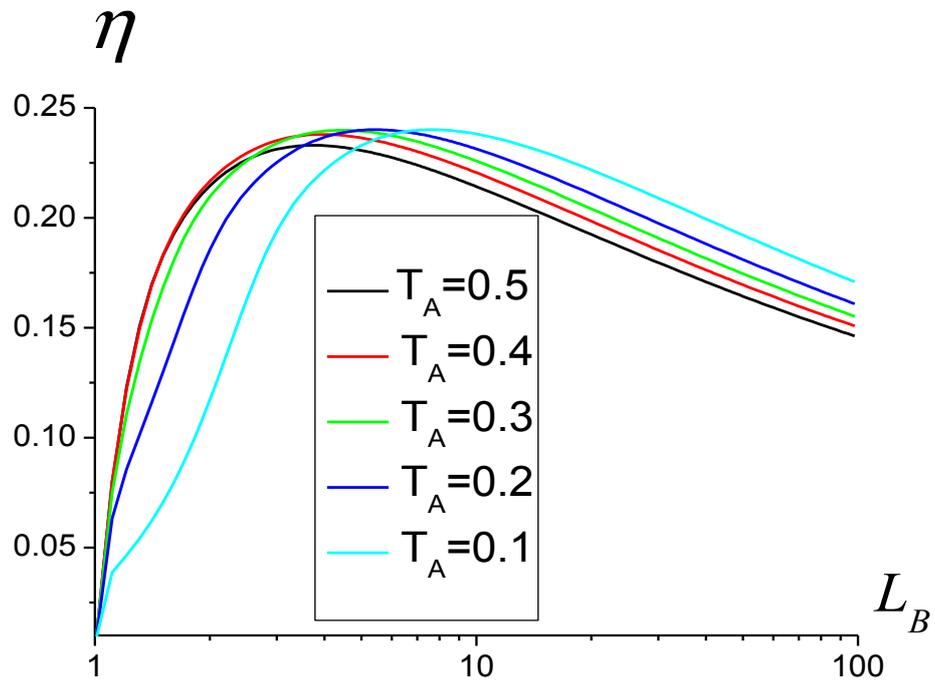

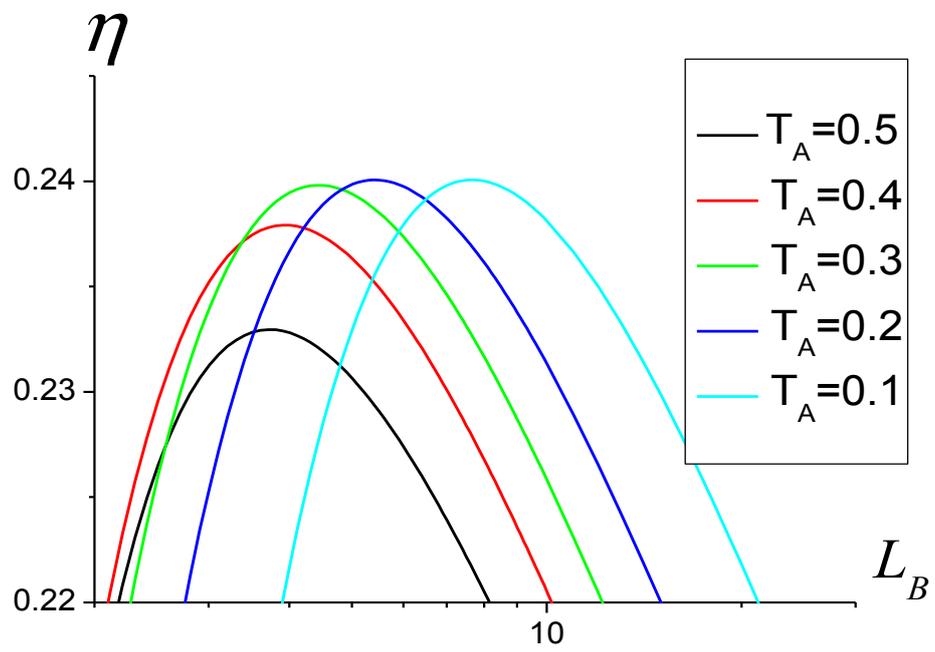

Fig. 3

Table 1

| $T_A$ | $L_B^*$ | $\eta_{max}$ |
|---|---|---|
| 2.0 | 4.61 | 0.131370 |
| 1.0 | 4.01 | 0.187521 |
| 0.5 | 3.71 | 0.232955 |
| 0.4 | 3.91 | 0.237916 |
| 0.3 | 4.51 | 0.239805 |
| 0.2 | 5.41 | 0.240062 |
| 0.1 | 7.71 | 0.240064 |